\documentclass[10pt, conference, compsocconf]{IEEEtran}

\usepackage{ifpdf}
\usepackage{amssymb,graphicx}
\usepackage{epsfig}
\usepackage{booktabs}
\usepackage[usenames]{color}
\usepackage{multirow}
\usepackage{rotating}

\usepackage{cite}
\usepackage{url}

\newcommand{\I}[1]{\textit{#1}}

\begin{document}
%
\title{Adaptive Mesh Refinement for Astrophysics Applications with ParalleX}


\author{
\IEEEauthorblockN{Matthew Anderson\IEEEauthorrefmark{1}, Maciej Brodowicz\IEEEauthorrefmark{2}, Hartmut Kaiser\IEEEauthorrefmark{2}\IEEEauthorrefmark{3}, Bryce Adelstein-Lelbach\IEEEauthorrefmark{2}, Thomas Sterling\IEEEauthorrefmark{1}}
\IEEEauthorblockA{\IEEEauthorrefmark{1}Center for Research in Extreme Scale Technology, Indiana University, Bloomington, IN}
\IEEEauthorblockA{\IEEEauthorrefmark{2}Center for Computation and Technology, Louisiana State University, Baton Rouge, LA}
\IEEEauthorblockA{\IEEEauthorrefmark{3}Department of Computer Science, Louisiana State University, Baton Rouge, LA}
\IEEEauthorblockA{andersmw@indiana.edu, maciek@cct.lsu.edu, hkaiser@cct.lsu.edu, blelbach@cct.lsu.edu, tron@indiana.edu}
}

\maketitle

\begin{abstract}
Several applications in astrophysics require adequately resolving many physical and temporal scales which vary 
over several orders of magnitude. Adaptive mesh refinement techniques address this problem effectively but 
often result in constrained strong scaling performance. The ParalleX execution model is an experimental execution 
model that aims to expose new forms of program parallelism and eliminate any global barriers present in a 
scaling-impaired application such as adaptive mesh refinement. We present 
two astrophysics applications using the ParalleX execution model: a 
tabulated equation of state component for neutron star
evolutions and a cosmology model evolution. Performance and 
strong scaling results from both simulations are presented. The tabulated 
equation of state data are distributed with transparent access over the nodes of the cluster. This allows 
seamless overlapping of computation with the latencies introduced by the remote access to the table. Because 
of the expected size increases to the equation of state table, this type of table partitioning for neutron star 
simulations is essential while the implementation is greatly simplified by ParalleX semantics.
\end{abstract}

\begin{IEEEkeywords}
Adaptive mesh refinement, astrophysics applications, ParalleX
\end{IEEEkeywords}

\IEEEpeerreviewmaketitle

%
%
\section{Introduction}

Among the most challenging problems in computational science is simulating the rich phenomenology 
of numerical relativity based astrophysical events like the collision of neutron stars combining 
electro-magnetics, extreme gravity, and neutrinos. Such complex simulations may provide the 
necessary signatures essential for the first arcane observations of gravitational waves from 
new instruments like the Laser Interferometric Gravitational Observatory (LIGO). To minimize 
unnecessary computational work, Adaptive Mesh Refinement (AMR) algorithms have been 
employed, greatly advancing the means of scientific discovery. However, like an increasing 
number of computational techniques in the emergent many-core era, true AMR-based problem 
performance can be severely limited by exhibiting strong scalability; in some cases taking 
weeks to compute a result but unable to effectively employ more than a few hundred 
processor cores using conventional practices such as the Communicating Sequential
Processes (CSP) execution model~\cite{csp} as reflected by the MPI based programming model~\cite{MPISpec}. 
A new computational strategy, replacing CSP, may be required to achieve dramatic 
increases in performance and continue to benefit from Moore's Law.

The ParalleX execution model~\cite{gao, scaling_impaired_apps, tabbal} is offered as a means of addressing these critical computational 
requirements. Performance for strong-scaled science codes like the HAD AMR~\cite{had_webpage} problem is 
determined by four factors: peak per unit capability, efficiency, scalability, and availability. 
The first is an innate property of a given hardware system and the last is related to issues 
of reliability (fault tolerance) and protection (i.e., security) both of which are outside the 
scope of the study being reported on. ParalleX is an experimental execution model developed 
to exploit runtime resource management and task scheduling to dramatically improve per locality 
(an equivalent of a traditional compute node)
efficiency and increase scaling of the number of localities that may be effectively employed. 
ParalleX is a synthesis of complementing semantic constructs delivering a dynamic adaptive 
framework for message-driven multi-threaded computing in a global address space context 
with constraint-based synchronization to exploit locality and manage asynchrony. The result 
is introspective runtime alignment of computing requirements and computing resources while 
managing asynchrony of operation across physically distributed resources. ParalleX has 
been first implemented in the form of the HPX runtime system~\cite{hpx_svn, amr1d} developed to support the 
semantics and mechanisms comprising ParalleX targeting conventional SMP and commodity 
cluster computing platforms. This experimental software package is developed to test the 
semantics of ParalleX, to measure the overhead costs of software implementation, and 
to provide a superior environment for extreme scale applications.

This paper discusses the results of employing the ParalleX execution model and applying 
the HPX runtime system to the HAD AMR simulation system for astrophysics applications 
with an emphasis on numerical relativity problems. These have proven to be exceptionally 
challenging and exemplify the growing set of strong-scaled algorithms that are failing to 
benefit from Moore's Law. This paper shows significant advances realized through improved 
efficiency with respect to static conventional methods and provides promising, although 
non-conclusive, results towards distributed scalability. The paper concludes with the 
implications of these results for future work.

%
%
\section{Adaptive Mesh Refinement for Astrophysics Applications}
Each of the two applications explored here each have several important 
physical scales.  Each of these scales must be adequately resolved in order
to properly understand the underlying dynamics.  In the cosmology application,
(see section~\ref{sec:cosmo}) the gradients at the domain wall and the subsequent
break require modeling several scales in the midst of exponential growth.  In 
the neutron star problem, the scales range from the internal dynamics of
the individual stars to the gravitational wave zone.  Strong scaling 
of such applications is typically poor~\cite{Anderson}.  For medium
sized applications, this often means that an AMR simulation would require 
weeks to months of runtime on a relatively small number of processors.  Due to the 
prevalence of such type simulations in astrophysics, we explore AMR in the
context of the ParalleX execution model.     

%
%
\section{The ParalleX Execution Model}

The ParalleX execution model~\cite{scaling_impaired_apps,tabbal,nbody,amr1d} 
was developed with the goal of specifying the next execution paradigm essential
to the full exploitation of future technology advances and computer architectures
in the near term as well as to guide co-design of computer architecture and
programming models in conjunction with supporting system software in the long term.
ParalleX is intended to catalyze innovation in system structure, operation,
and applications to realize practical Exascale processing capability by the end of this decade.

The development of ParalleX is motivated by two challenges we face when 
developing certain classes of applications with conventional models.
Scaling-impaired applications, such as the described astrophysics adaptive mesh
refinement codes, are usually unable to effectively exploit a relatively small 
number of cores in a multi-core system. Such applications will likely also be
unable to exploit future Exascale computing systems. Four factors are
inhibiting their scalability:

\begin{itemize}
\item \I{Starvation}, the unavailability of useful work either globally or locally.
\item \I{Latency}, delays due to remote accesses or service requests.
\item \I{Overhead}, the critical time and work required to manage parallel
resources and concurrent tasks which would not be required for pure sequential
execution.
\item \I{Waiting for contention resolution}, delays due to conflicts for shared physical 
or logical resources.
\end{itemize}

The ParalleX execution model strives to overcome these 
limitations through four principal properties:

\begin{itemize}
\item Exposure of intrinsic parallelism, especially from meta-data, 
to meet the concurrency needs of scalability by systems in the next decade.
\item Intrinsic system-wide latency hiding for superior time and power efficiency.
\item Dynamic adaptive resource management for greater efficiency by exploiting runtime information.
\item A global name space to reduce the semantic gap between application requirements and
system functionality both to enhance programmability and to improve overall system utilization and efficiency.
\end{itemize}

While ParalleX incorporates many useful concepts developed
elsewhere, some extending back as far as three decades, it
constitutes a new synthesis of these as well as innovative ideas in a
novel schema that is distinct from conventional practices and that
exhibits the necessary properties identified above to increase
application and system scalability.
The form and function of the current ParalleX model consist of six key concepts
or management principles: ParalleX Processes (PX-processes), the Active Global Address Space (AGAS),
threads (PX-threads) and their management, parcel transport and their management, Local Control Objects (LCOs),
and percolation~\cite{jacquet_percolation_03}. With the exception of PX-processes and percolation, all have been incorporated
in a C++ prototype runtime implementation of ParalleX called HPX (High Performance ParalleX)~\cite{hpx_svn, amr1d}.  
We will highlight these concepts throughout this paper wherever they relate to our results.

%
%
\section{The HPX Runtime System}

\begin{figure*} 
  \includegraphics[width=0.99\linewidth,height=0.4\textheight]{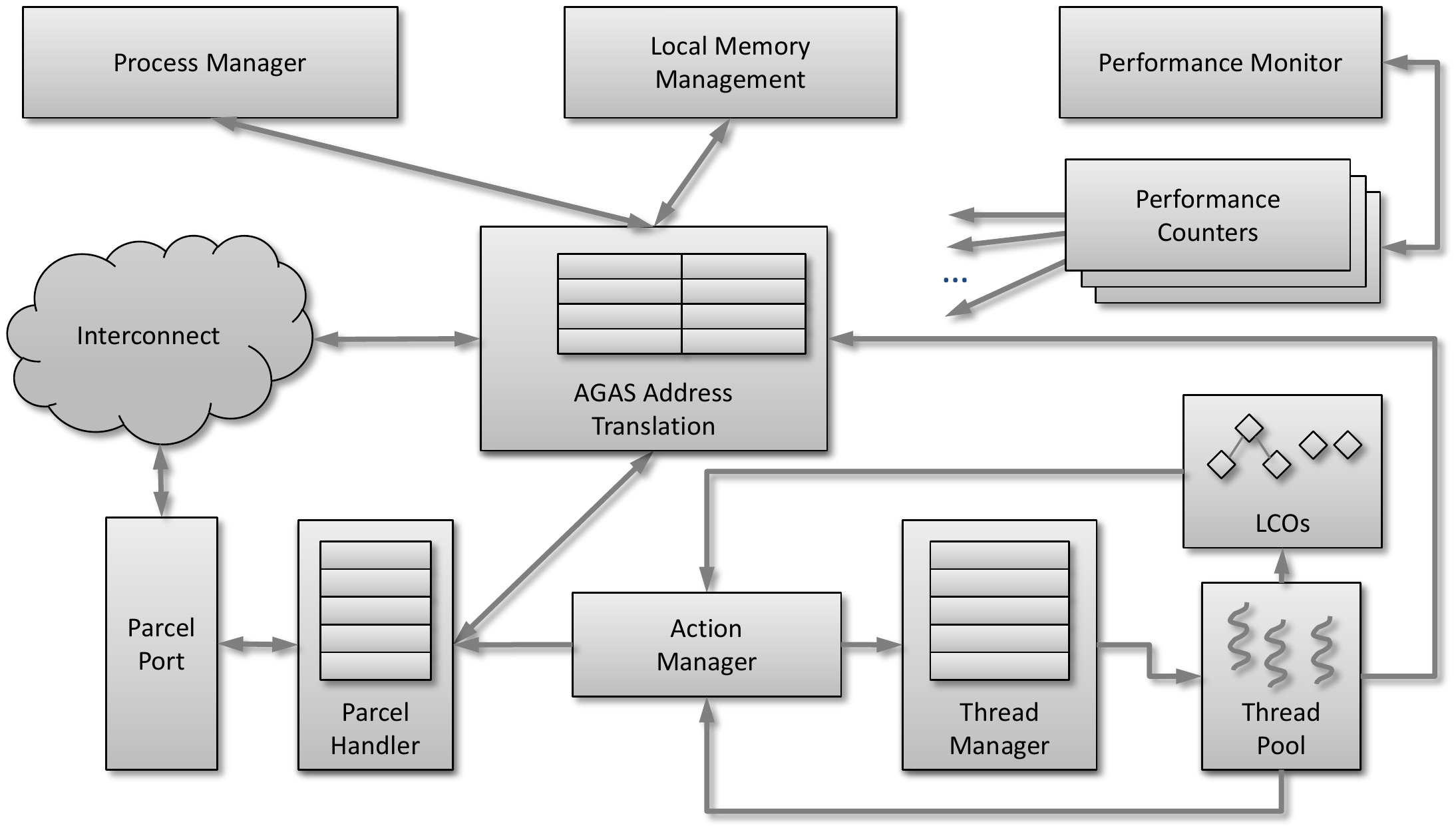}
  \caption{\small{Modular structure of HPX implementation. HPX implements the supporting
     functionality for all of the elements needed for the ParalleX
     model: AGAS (active global address space), parcel port and
     parcel handlers, HPX-threads and thread manager, ParalleX processes, LCOs (local control objects),
     performance counters enabling dynamic and intrinsic system and load
    estimates, and the means of integrating application specific components.}
  }
\label{fig:hpxarch}
\end{figure*}

A walkthrough description of the HPX architecture is found in Figure~\ref{fig:hpxarch}. An incoming parcel 
(delivered over the interconnect) is received by the parcel port. One or more 
parcel handlers are connected to a single parcel port, optionally allowing to distinguish different 
parts of the system as the parcel's final destination. An example for such different destinations 
is to have both normal cores and special hardware (such as a GPGPU) in the same locality. The 
main task of the parcel handler is to buffer incoming parcels for the action manager. The action 
manager decodes the parcel, which contains an action. An action is either a global function
call or a method call on a globally addressable object. The action manager creates a PX-thread
based on the encoded information.

All PX-threads are managed by the thread manager, which schedules their execution on one 
of the OS-threads allocated to it. Usually HPX creates one OS-thread for each available core. Several scheduling 
policies have been implemented for the thread manager, such as the global queue scheduler, where all
cores pull their work from a single, global queue, or the local queue scheduler, where each core
pulls its work from a separate queue. The latter supports work stealing for better
load balancing. In the local scheduler, a queue is created for each of the OS-threads
dedicated to the thread manager. These queues are placed in a contiguous block of local
memory. When an OS-thread is searching for work, it first checks its
own queue. If there is no work there, the OS-thread begins to
steal work by searching for work in other queues.

If a possibly remote action has to be executed by a PX-threads, the action manager 
queries the global address space (AGAS) to determine whether the target of the action is local or
remote to the locality that the PX-thread is running on. If the target happens to be local, a new
PX-thread is created and passed to the thread manager. This thread encapsulates the work (function)
and the corresponding arguments for that action. If the target is remote, the action manager
creates a parcel encoding the action (i.e. the function and its arguments). This parcel is handed
to the parcel handler, which makes sure that it gets sent over the interconnect.

The Active Global Address Space (AGAS) provides global address resolution services
that are used by the parcel port and the action manager. AGAS addresses are 128bit
unique global identifiers (GIDs). AGAS maps these global identifiers to local addresses,
and additionally provides symbolic mappings from strings to GIDs. The local addresses
that GIDs are bound to are typed, providing a degree of protection from type errors.
Any object that has been registered with a GID in AGAS is addressable from all localities
in an instance of the HPX runtime. AGAS also provides a powerful reference counting
system which implements global garbage collection.

Lightweight Control Objects (LCOs) are the synchronization primitives upon which
HPX applications are built. LCOs provide a means of controlling parallelization
and synchronization of PX-threads. Semaphores, mutexes and condition
variables~\cite{mutex} are all available in HPX as LCOs. Futures~\cite{future1}
are another type of LCO provided by HPX, and are discussed in greater detail
later in this paper. 

Local memory management, performance counters (a generic monitoring framework),
LCOs and AGAS are all implemented on top of an underlying component framework.
Components are the main building blocks of remotable actions and can encapsulate
arbitrary, possibly application specific functionality. Actions expose the
functionality of a component. An action can be invoked on a component instance
using only its GID, which allows any locality to
invoke the exposed methods of a component. In the case of the aforementioned components,
the HPX runtime system implements its own functionality in terms of this component framework.
Typically, any application written using HPX extends the set of existing components based on
its requirements.

%
%
\section{Using Shen Equation of State Tables}

The Shen equation of state (EOS) tables of nuclear matter at finite temperature and density 
with various electron fractions within the relativistic mean field (RMF) 
theory~\cite{1998NuPhA.637..435S} are a set of three dimensional data
arrays enabling high precision interpolation of 19 relevant parameters required for 
neutron star simulations.  While these tables are currently relatively 
small in size (about 300\,MBytes), it is expected that over the next year a 
new set of tables ensuring higher resolution will be published. The size of the new tables is expected
to be in the range of several GBytes. This will prevent loading the whole data set into 
main memory on each locality. In conventional, MPI based applications the full tables 
would have to be either loaded into each MPI process or a distributed partitioning
scheme would have to be implemented. Both options are either not viable or difficult 
to implement. 

\subsection{The Overhead of Eager Futures}
\label{subsec:futures}

Many HPX applications, including the astrophysics applications detailed here,
utilize Futures for ease of parallelization and synchronization. In HPX, Futures
are implemented as two types of LCOs (Eager Futures and Lazy Futures). The
astrophysics applications detailed in this paper make extensive use of Eager
Futures. For this reason, the overheads of these constructs is a large factor in
the total overhead of the HPX runtime in our code. In this subsection, we give a
description of Futures, outline a performance test for measuring the overhead of
Futures, and present the results of the test. 

\begin{figure} 
  \includegraphics[width=0.99\linewidth]{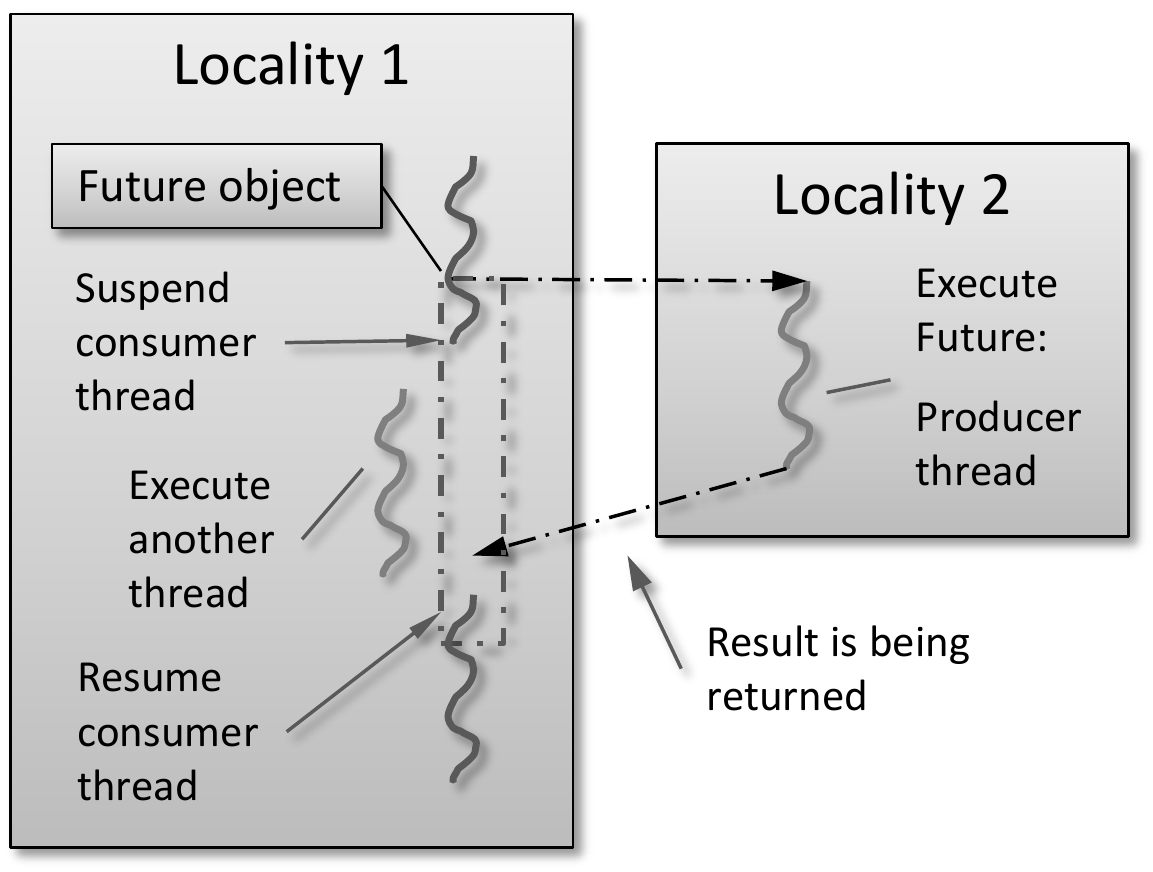}
  \caption{\small{Schematic of a Future execution. At the point of creation of the 
     Future, its encapsulated execution is started. The consumer thread is suspended only 
     if the result of executing the Future  has not returned yet. In this case the core is free to 
     execute some other work (here 'another thread') in the meantime. If the result is available,
     the consumer thread continues without interruption to complete execution. The producer 
     thread may be executed on the same locality as the consumer thread or on a different 
     locality, depending on whether the target data is local or not.} 
  }
\label{fig:future_schematics}
\end{figure}

As shown in Figure~\ref{fig:future_schematics}, a Future encapsulates a delayed 
computation. It acts as a proxy for a result initially not known, most of the
time because the computation of the result has not completed yet. The
Future synchronizes the access of this value by optionally suspending
PX-threads requesting the result until the value is available. When a Future is
created, it spawns a new PX-thread (either remotely with a parcel or locally
by placing it into the thread queue) which, when run, will execute the
action associated with the Future. The arguments of the action are bound when
the Future is created. 
Once the action has finished executing, a write operation is performed on the
Future. The write operation marks the Future as completed, and
optionally stores data returned by the action. 

When the result of the delayed
computation is needed, a read operation is performed on the Future. If the
Future's action hasn't completed when a read operation is performed on it, the
reader PX-thread is suspended until the Future is ready. 

\begin{figure} 
  \includegraphics[width=0.99\linewidth]{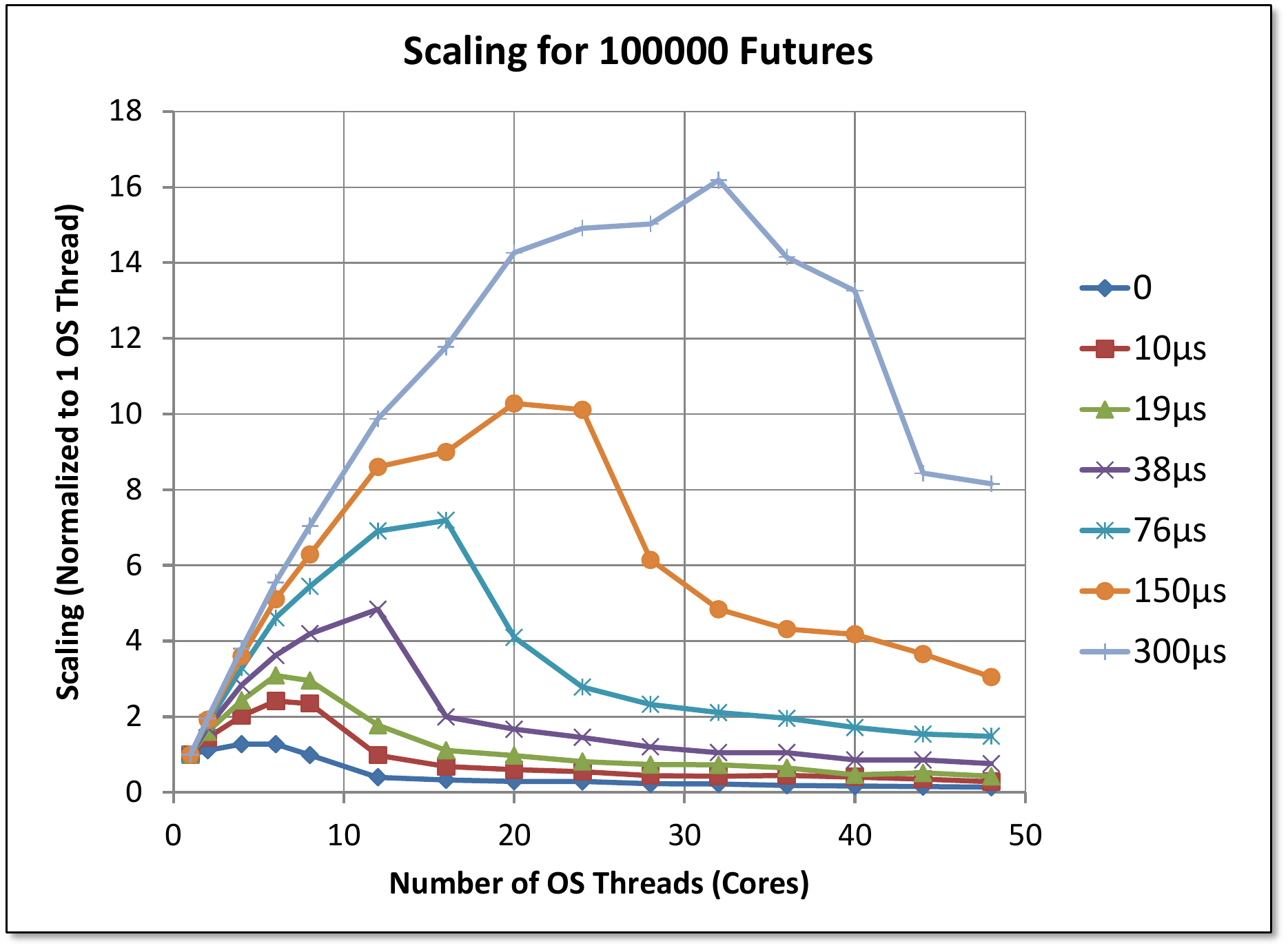}
  \caption{\small{Results of the Eager Future overhead benchmark. In each
    test, 100,000 Futures were invoked, with varying workloads. Each
    data set shows the total execution time for a particular workload. The
    smallest workload was an empty action, and shows the baseline overhead
    of HPX's Eager Future implementation. The largest workload was 300
    microseconds.} 
  }
\label{fig:eager_future_overhead}
\end{figure}

Our benchmark for Future overhead created a fixed number of Futures,
each of which had a fixed workload. Then, asynchronous read operations were
performed on the Futures until all of the Futures had completed. A high
resolution timer measured the wall-time of the aforementioned operations. The
test was run on an 8-socket HP ProLiant DL785 (each socket sports a 6-core AMD
Opteron 8431) with 96 Gbytes of RAM (533 MHz DDR2). Trials were done with
varying workloads and OS-threads. 5 runs were done for each combination of
the parameters, and the results were averaged to produce a final dataset. The
numbers are presented in Figure~\ref{fig:eager_future_overhead}.

On the locality we used for this benchmark, the amortized overhead of an 
Eager Future is approximately \emph{40 microseconds}. This number was extrapolated
from the data presented in Figure~\ref{fig:eager_future_overhead}. We multiplied
the workload by the number of Futures used in each run, and then subtracted
that from the average wall-time of the trial. We divided that number by the 
number of  Futures invoked in the trial to get the overhead per Future for
each set of parameters. 

\begin{eqnarray*}
& \mbox{\textit{overhead}} = \frac{\mbox{\textit{avg. wall-time}}-(\mbox{\textit{workload}}*\mbox{\textit{futures invoked}})}{\mbox{\textit{futures invoked}}}
\end{eqnarray*}
 
The scaling results in Figure~\ref{fig:eager_future_overhead} call for some
explanation. The parabolic curves are formed primarily by contention in the
thread queue scheduler. As the number of OS-threads is increased, the contention
on the thread queue scheduler also increases, due to a higher number of
concurrent searches for available work. This increased contention occurs in both
global queue schedulers (where all OS-threads poll the same work queue, and must
obtain exclusive access to said queue for some period of time) and to a lesser
degree in local queue schedulers (where work stealing occurs, which causes queue
contention, albeit to a lesser degree than in the global queue scheduler). As
we increase the workload in each Future, OS-threads spend more time
executing the workloads and less time searching for more work. This decreases 
contention on the queues. Adding a new OS-thread is beneficial as long as the
contention overhead that it causes is not greater than the parallel speedup that
it provides. 

\subsection{The Overhead of the ShenEOS Table Partitioning}
\label{subsec:sheneos}

\begin{figure} 
  \includegraphics[width=0.99\linewidth]{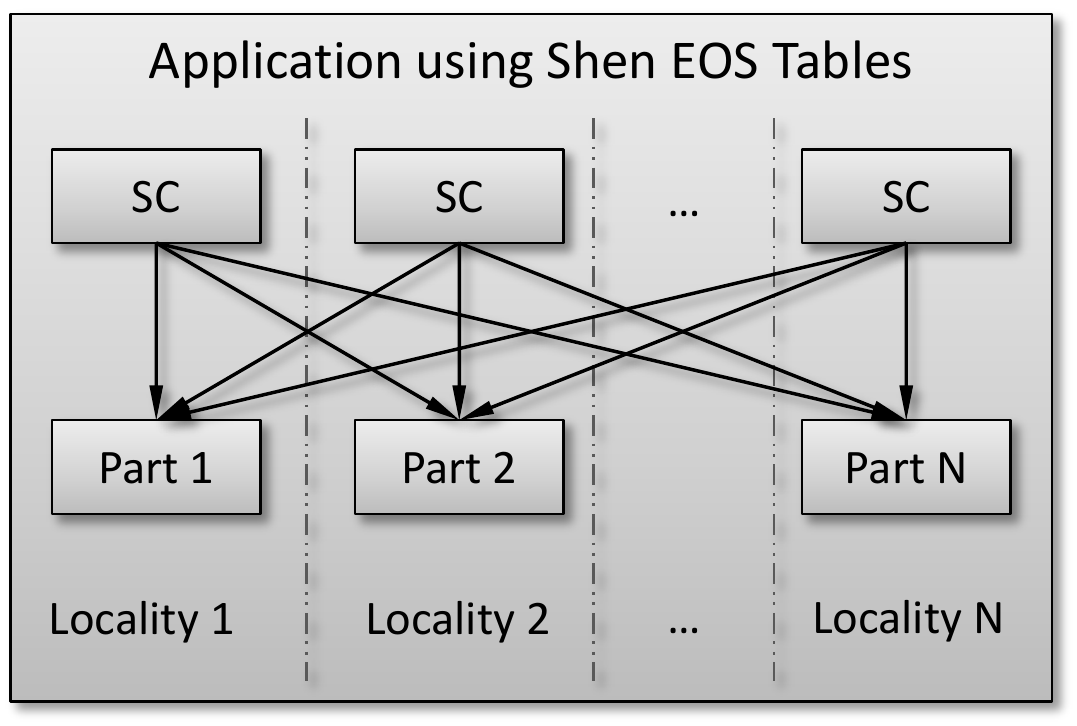}
  \caption{\small{Schematic of an application using the distributed partitioned Shen EOS (equation of state) 
    tables. Each locality has a ShenEOS client side object (SC) allowing to transparently access all of the table data.
    At the same time the Shen EOS table data is partitioned into equally sized chunks, each of which is loaded into
    the main memory of one of the localities (Part 1~...~Part N) thus lessening the required memory footprint
    for each of the localities.}
  }
\label{fig:sheneos}
\end{figure}

We created an HPX component encapsulating the non-overlapping partitioning and 
distribution of the Shen EOS tables to all available localities, thus reducing the required 
memory footprint per locality~\cite{sheneos_svn}. A special client side object ensures the transparent 
dispatching of interpolation requests to the appropriate partition corresponding to 
the locality holding the required part of the tables (see Figure~\ref{fig:sheneos}).
The client side object exposes a simple API for easy programmability.

The second part of this section describes the setup and results of the
measurements we performed in order to estimate the overheads introduced
by distributing the Shen EOS tables across all localities. To evaluate the
scalability and associated overheads of the distributed implementation
of the Shen EOS tables, a number of tests has been
performed, all of them with a fixed number of total data accesses
(measuring strong scaling). The tests have been run on a different
number of localities and with varying numbers of OS-threads per locality.
The current HPX implementation supports only a centralized AGAS server
that may be invoked in two configurations: either as a standalone task
on a dedicated locality or as a part of one of the user application tasks.
Our tests used a standalone AGAS server, firstly to avoid interfering with
the user workload and secondly to eliminate the
generation of asymmetric AGAS traffic on localities hosting data
tables. Unlike the client applications, the AGAS server used a
fixed number of OS-threads throughout the testing to ensure that
sufficient processing resources are available to the incoming
resolution requests.

The tests were performed on a small heterogeneous cluster. The cluster consists
of 18 localities (excluding the head node) connected by Gigabit Ethernet network. Two of
the machines are 8-socket HP ProLiant DL785s, with 6-core AMD Opteron 8431s and 96
Gbytes of RAM (533 MHz DDR2). The other 16 localities are single-socket HP ProLiant DL120s,
with Intel Xeon X3430s and 4 Gbytes of RAM (1332 MHz DDR3). All machines run
x86-64 Debian Linux. Torque PBS was used to run multi-locality tests. 

\begin{figure} 
  \includegraphics[width=0.99\linewidth]{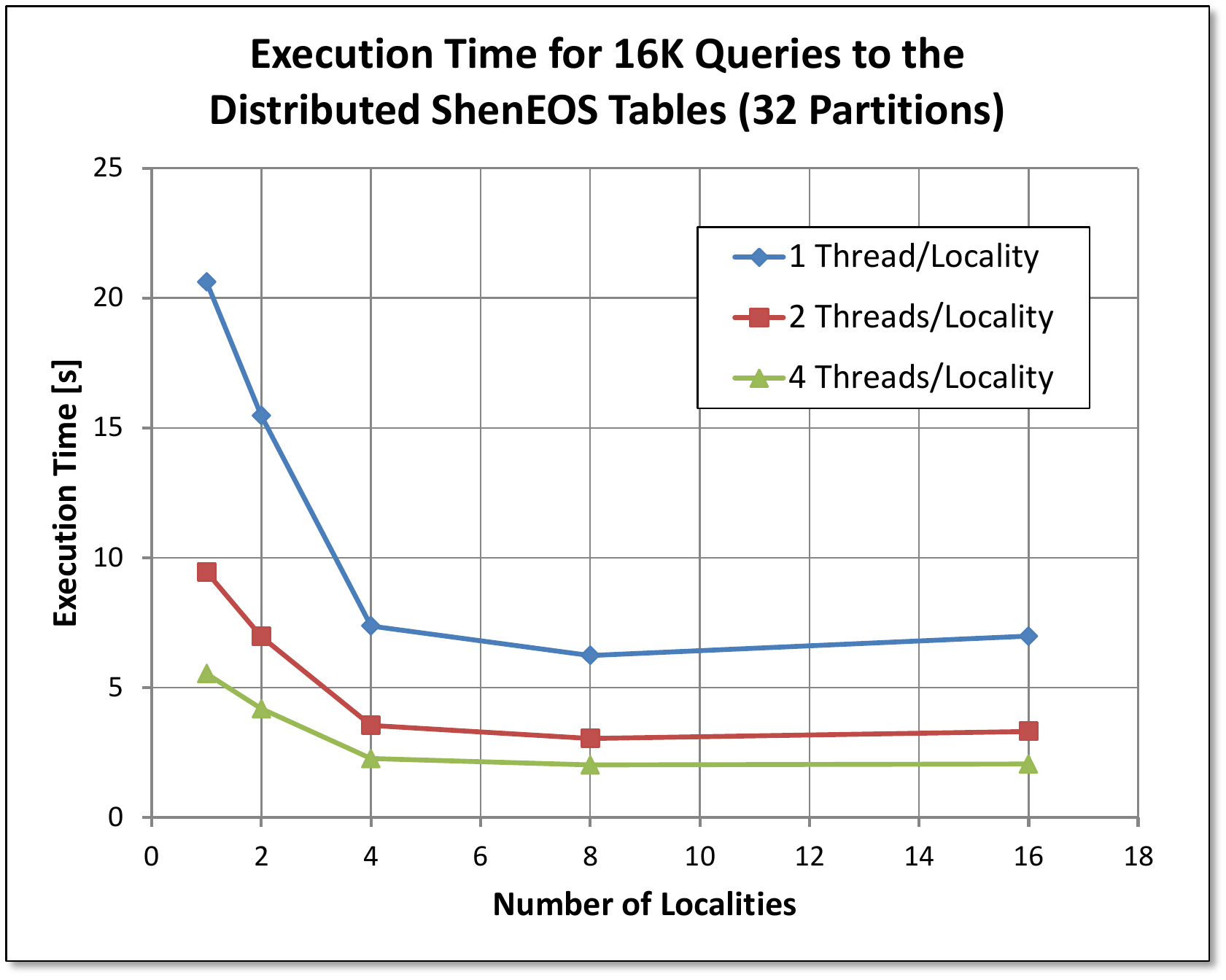}
  \caption{\small{Scaling of the execution time for 16K data interpolation
    queries to the Shen EOS tables distributed across 32 partitions measured 
   for different number of localities and varying number of OS-threads per locality.}
  }
\label{fig:sheneos_execution_time}
\end{figure}

Figure~\ref{fig:sheneos_execution_time} shows the execution times
collected for the data access phase with a special test application executed 
on up to 16 localities and 1, 2, and 4 OS-threads per locality. The total number of
distributed partitions was fixed at 32 to preserve the AGAS traffic
pattern when run on a different number of localities; all partitions
were uniformly distributed across the test localities. The number of 
queries to the distributed ShenEOS partitions was set to a fixed 
number of 16K. Each of these queries created a Eager Future encapsulating
the whole operation of sending the request to the remote partition,
schedule and execute a PX-thread, perform the interpolation based on
the supplied arguments for the ShenEOS data, sending back the resulting
values to the requesting PX-thread, and resuming the PX-thread which was
suspended by the Future in order to wait for the results to come back.


The graph demonstrates that the overhead of distributed table
implementation does not increase significantly over the entire range
of available localities.  While the scaling is much better when the
number of localities remains small (up to 4), the overall time
required to service the full 16K data lookup requests remains roughly
constant, which proves the quality of the implementation. 
The test application itself does not execute any work besides 
querying the distributed tables, which does not leave any room
to overlap the significant network traffic generated with useful
computation. This causes the scaling to flatten out beyond
8 localities. Using the distributed tables in real applications doing much more
work will allow to further amortize the introduced network overheads. 
The results also imply that a single AGAS server is capable of servicing at least 16
client localities, especially considering the intensity of request
traffic over Ethernet interconnect deployed in out testbed.  We plan
to further evaluate this aspect of distributed table implementation
using faster interconnect networks, such as Infiniband.

%
%
\section{Cosmology Application}
\label{sec:cosmo}
In cosmology, inflation is driven by the inflaton field, which is a
scalar field with a nonlinear potential.  Quantum fluctuations in the
inflaton field will result in some regions of space-time expanding
more rapidly than other regions.  The boundary between these regions
is the domain wall, where the inflaton field changes from one vacuum
value to another over a small region.  The cosmology code models
the dynamics of the domain wall.  The scalar field is the inflaton
field with a $\psi^4$ potential.  The initial data are a kink, modeling
the transition from one vacuum state to another.  This application aims to study the
stability and dynamics of the domain wall during the cosmological
expansion.  The domain wall eventually has to break because the
region that is expanding more rapidly does so exponentially.
Depending on how the domain walls break or snap, there may be
observational signatures in the cosmic microwave background.

\begin{figure} 
  \includegraphics[width=0.99\linewidth]{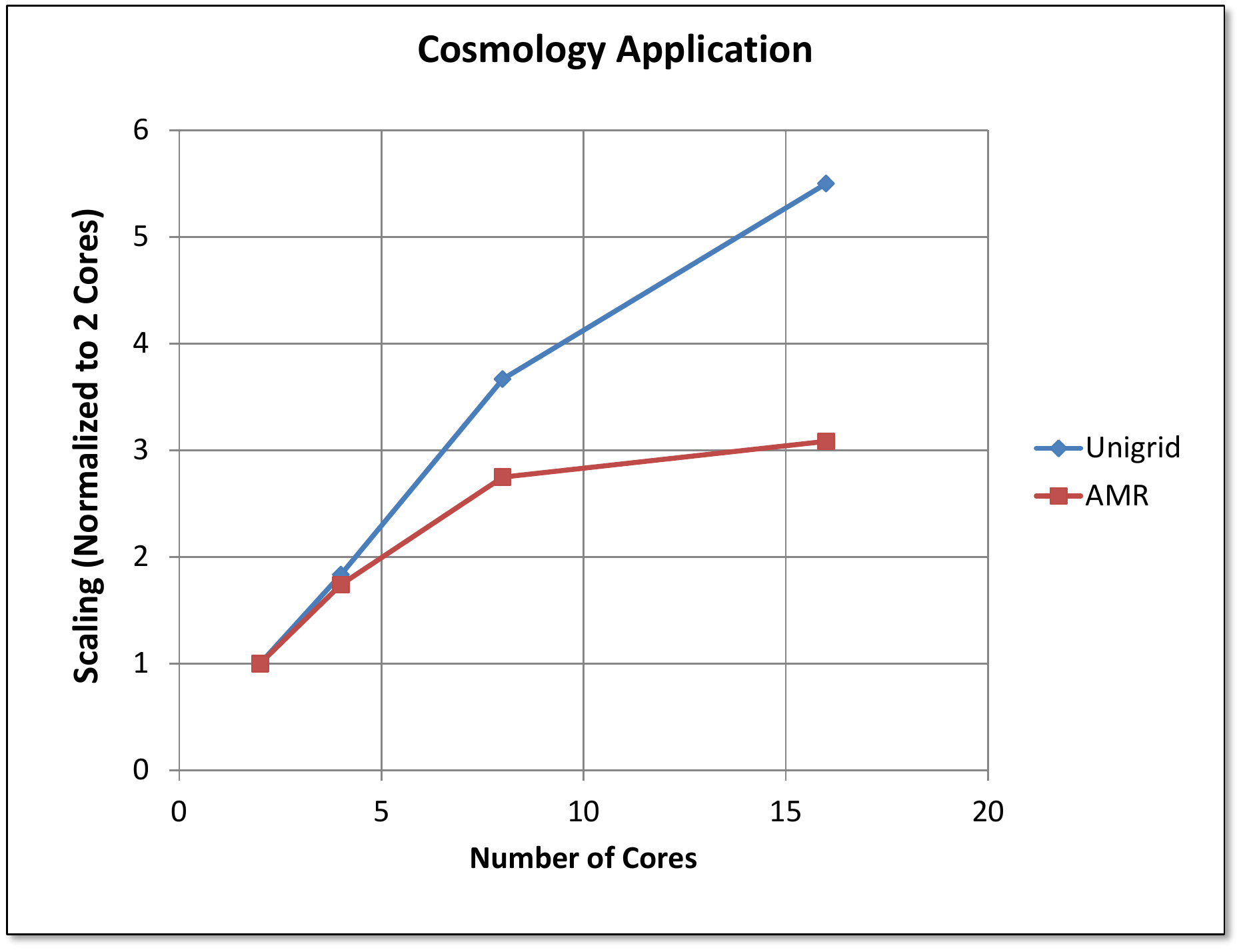}
  \caption{\small{Strong scaling results for the cosmology application on
a distributed quad-core Intel Nehalem based cluster.  The interconnect between
localities was Gigabit Ethernet.}
  }
\label{fig:amr}
\end{figure}

The dynamics of the domain wall during expansion encompass many
temporal and spatial scales making it an excellent candidate for AMR
simulation.   The system of equations has nine variables ($\phi$,$\Pi$,$\chi$,$a$,$f$,$g$,$b$,$q$,$r$)
and two independent variables ($t$ and $z$):
\begin{eqnarray*}
\frac{\partial \phi}{\partial t} &=& \Pi \\
\frac{\partial \Pi}{\partial t} &=& \Pi \left(f/a + q/b\right)  + \left(3 a g / b^2 - a^2 r/b^3 \right)\chi \\
                                & &  + \left(a/b\right)^2 \frac{\partial \chi}{\partial z}  
                                  - a^2 \lambda \phi \left(\phi^2 - v^2\right)\\
\frac{\partial \chi}{\partial t} &=& \frac{\partial \Pi}{\partial z}\\
\frac{\partial a}{\partial t} &=& f \\
\frac{\partial f}{\partial t} &=& a ( -(f q)/(a b) + 2 g^2/b^2 - a g r/b^3 + a \frac{\partial g}{\partial z}/b^2\\
                              & &  + a^2 \lambda \left(\phi^2 - v^2\right)/4 )\\
\frac{\partial g}{\partial t} &=& \frac{\partial f}{\partial z}\\
\frac{\partial b}{\partial t} &=& q \\
\frac{\partial q}{\partial t} &=& b ( -(f q)/(a b) -3 a g r/b^3 + 3 a \frac{\partial g}{\partial z}/b^2 \\ 
                              & & + a^2 \lambda \left(\phi^2 - v^2\right)/4 )\\
\frac{\partial r}{\partial t} &=& \frac{ \partial q}{\partial z},
\end{eqnarray*}
where $\lambda = 1$, $v = 0.1$, and the boundary condition is periodic.  
The system is evolved with second order differencing in space and third order Runge-Kutta integration in time.
The AMR algorithm is Berger-Oliger~\cite{Berger} but uses tapering~\cite{Lehner:2005vc} to reduce noise at
coarse-fine interfaces.  The algorithm is implemented using dataflow LCOs, thereby replacing global barriers
with constrained synchronization.  Performance results on a quad-core Intel Nehalem (2.8 GHz) cluster with 
1333 MHz DDR3 memory are detailed in Figure~\ref{fig:amr}.     

%
%
\section{Conclusion}

In this paper we have presented performance and scaling measurements for two key astrophysics 
applications built with 
the High Performance ParalleX C++ library (HPX). We have presented overhead measurements for one of the building blocks of these applications (Futures) and
an application specific distributed partitioning component (ShenEOS). The results of these benchmarks show sufficiently good
strong-scaling behavior to conclude that these components will not significantly impede 
full neutron star applications when fully integrated therein.
The amortized overhead for creating, using and deleting one Future object has been 
establish to be approximately 40 microseconds. 

We have also presented distributed AMR scaling results for a cosmology application.
Astrophysical applications using AMR, which are known to expose 
unsatisfactory strong scaling behavior when implemented using conventional (MPI based)
techniques, are not only easier to implement when built on top of HPX, but also exhibit promising
strong scaling characteristics in distributed runs.

Future work will be focusing on reducing the overheads introduced by HPX and 
developing currently unimplemented parts of the ParalleX execution model (such
as PX-processes and percolation). Reducing the overhead of Futures and
the HPX scheduling subsystem will improve HPX's scaling potential. To enable
HPX applications to utilize thousands of localities for a single application,
the overheads of the contention on the centralized AGAS server must be reduced
by distributing the AGAS subsystem across multiple localities. 

%
%
\section*{Acknowledgments}

We acknowledge support comes from NSF grants 1048019, 1029161, and 1117470 to Louisiana State University.


\bibliographystyle{IEEEtran}
\bibliography{IEEEabrv,pxBib}

\begin{thebibliography}{10}
\providecommand{\url}[1]{#1}
\csname url@samestyle\endcsname
\providecommand{\newblock}{\relax}
\providecommand{\bibinfo}[2]{#2}
\providecommand{\BIBentrySTDinterwordspacing}{\spaceskip=0pt\relax}
\providecommand{\BIBentryALTinterwordstretchfactor}{4}
\providecommand{\BIBentryALTinterwordspacing}{\spaceskip=\fontdimen2\font plus
\BIBentryALTinterwordstretchfactor\fontdimen3\font minus
  \fontdimen4\font\relax}
\providecommand{\BIBforeignlanguage}[2]{{%
\expandafter\ifx\csname l@#1\endcsname\relax
\typeout{** WARNING: IEEEtran.bst: No hyphenation pattern has been}%
\typeout{** loaded for the language `#1'. Using the pattern for}%
\typeout{** the default language instead.}%
\else
\language=\csname l@#1\endcsname
\fi
#2}}
\providecommand{\BIBdecl}{\relax}
\BIBdecl

\bibitem{csp}
S.~D. Brookes, C.~A.~R. Hoare, and A.~W. Roscoe, ``A theory of communicating
  sequential processes,'' \emph{J. ACM}, vol.~31, no.~3, pp. 560--599, 1984.

\bibitem{MPISpec}
{Message Passing Interface Forum}, \emph{MPI: A Message-Passing Interface
  Standard, Version 2.2}.\hskip 1em plus 0.5em minus 0.4em\relax Stuttgart,
  Germany: {High Performance Computing Center Stuttgart (HLRS)}, September
  2009.

\bibitem{gao}
G.~Gao, T.~Sterling, R.~Stevens, M.~Hereld, and W.~Zhu, ``{Parallex: A study of
  a new parallel computation model},'' in \emph{{Parallel and Distributed
  Processing Symposium, 2007. IPDPS 2007. IEEE International}}, 2007, pp. 1--6.

\bibitem{scaling_impaired_apps}
H.~Kaiser, M.~Brodowicz, and T.~Sterling, ``{ParalleX}: An advanced parallel
  execution model for scaling-impaired applications,'' in \emph{Parallel
  Processing Workshops}.\hskip 1em plus 0.5em minus 0.4em\relax Los Alamitos,
  CA, USA: IEEE Computer Society, 2009, pp. 394--401.

\bibitem{tabbal}
A.~Tabbal, M.~Anderson, M.~Brodowicz, H.~Kaiser, and T.~Sterling, ``Preliminary
  design examination of the {ParalleX} system from a software and hardware
  perspective,'' \emph{SIGMETRICS Performance Evaluation Review}, vol.~38,
  p.~4, Mar 2011.

\bibitem{had_webpage}
\BIBentryALTinterwordspacing
``{\sc had} home page.'' [Online]. Available: \url{http://www.had.liu.edu/}
\BIBentrySTDinterwordspacing

\bibitem{hpx_svn}
\BIBentryALTinterwordspacing
H.~Kaiser, B.~Adelstein-Lelbach \emph{et~al.}, ``{HPX SVN repository},'' 2011,
  available under a BSD-style open source license. Contact gopx@cct.lsu.edu for
  repository access. [Online]. Available:
  \url{https://svn.cct.lsu.edu/repos/projects/parallex/trunk/hpx}
\BIBentrySTDinterwordspacing

\bibitem{amr1d}
\BIBentryALTinterwordspacing
M.~Anderson, M.~Brodowicz, H.~Kaiser, and T.~Sterling, ``An application driven
  analysis of the {ParalleX} execution model,'' 2011. [Online]. Available:
  \url{http://arxiv.org/abs/1109.5201}
\BIBentrySTDinterwordspacing

\bibitem{Anderson}
M.~Anderson, E.~Hirschmann, S.~Liebling, and D.~Neilsen, ``{MHD} with adaptive
  mesh refinement,'' \emph{Class. {Q}uant. {G}rav.}, vol.~23, pp. 6503--6524,
  2006.

\bibitem{nbody}
\BIBentryALTinterwordspacing
C.~Dekate, H.~Kaiser, M.~Anderson, B.~Adelstein-Lelbach, and T.~Sterling,
  ``Improving the scalability of parallel n-body applications with an event
  driven constraint based execution model,'' 2011. [Online]. Available:
  \url{http://arxiv.org/abs/1109.5190}
\BIBentrySTDinterwordspacing

\bibitem{jacquet_percolation_03}
A.~Jacquet, V.~Janot, G.~R. Gao, C.~Leung, R.~Govindarajan, and T.~Sterling,
  ``An executable analytical perfromance evaluation approach for early
  performance prediction,'' in \emph{In Proc. of the Workshop on Massively
  Parallel Processing held in conjunction with Intl. Parallel and Distributed
  Processing Symposium ({IPDPS}-03)}, Nice, France, April 2003.

\bibitem{mutex}
P.~J. Courtois, F.~Heymans, and D.~L. Parnas, ``Concurrent control with
  ``readers'' and ``writers'','' \emph{Commun. ACM}, vol.~14, no.~10, pp.
  667--668, 1971.

\bibitem{future1}
\BIBentryALTinterwordspacing
H.~C. Baker and C.~Hewitt, ``The incremental garbage collection of processes,''
  in \emph{SIGART Bull.}\hskip 1em plus 0.5em minus 0.4em\relax New York, NY,
  USA: ACM, August 1977, pp. 55--59. [Online]. Available:
  \url{http://doi.acm.org/10.1145/872736.806932}
\BIBentrySTDinterwordspacing

\bibitem{1998NuPhA.637..435S}
H.~{Shen}, H.~{Toki}, K.~{Oyamatsu}, and K.~{Sumiyoshi}, ``{Relativistic
  equation of state of nuclear matter for supernova and neutron star},''
  \emph{Nuclear Physics A}, vol. 637, pp. 435--450, Jul. 1998.

\bibitem{sheneos_svn}
\BIBentryALTinterwordspacing
H.~Kaiser, ``{HPX Shen EOS SVN repository},'' 2011, available under a BSD-style
  open source license. Contact gopx@cct.lsu.edu for repository access.
  [Online]. Available:
  \url{https://svn.cct.lsu.edu/repos/projects/parallex/trunk/hpx/examples/shen%
eos}
\BIBentrySTDinterwordspacing

\bibitem{Berger}
M.~J. Berger and J.~Oliger, ``Adaptive mesh refinement for hyperbolic partial
  differential equations,'' \emph{J. Comp. Phys.}, vol.~53, p. 484, 1984.

\bibitem{Lehner:2005vc}
L.~Lehner, S.~L. Liebling, and O.~Reula, ``{AMR}, stability and higher
  accuracy,'' \emph{Class. Quant. Grav.}, vol.~23, pp. S421--S446, 2006.

\end{thebibliography}

\end{document}